\documentclass[pdftex,twocolumn,epjc3]{svjour3}         

\RequirePackage[T1]{fontenc}

\smartqed  

\RequirePackage{graphicx}
\RequirePackage{mathptmx}      
\RequirePackage{flushend}
\RequirePackage[numbers,sort&compress]{natbib}
\RequirePackage[colorlinks,citecolor=blue,urlcolor=blue,linkcolor=blue]{hyperref}

\usepackage{graphicx}
\graphicspath{{fig/}}
\usepackage{siunitx}
\usepackage{hyperref}
\usepackage{epsfig}
\usepackage{booktabs}
\usepackage{multirow}
\usepackage{slashbox} 
\usepackage{lipsum}
\usepackage{mathtools}
\usepackage{cuted}
\hypersetup{hidelinks,colorlinks=false,breaklinks=true,urlcolor=ocre}

\journalname{Eur. Phys. J. C}

\begin{document}

\title{The Impact of the Charge Barrier Height on Germanium (Ge) Detectors with Amorphous-Ge Contacts for Light Dark Matter Searches}


\author{W.-Z. Wei\thanksref{addr1}
        \and
        R. Panth\thanksref{addr1}, 
        J. Liu\thanksref{addr1},
        H. Mei\thanksref{addr1}, 
        D.-M. Mei\thanksref{e1,addr1} and
        G.-J. Wang\thanksref{addr1}
        }

\thankstext{e1}{Corresponding Author: Dongming.Mei@usd.edu}

\institute{Department of Physics, The University of South Dakota, Vermillion, South Dakota 57069\label{addr1}}

\date{Received: date / Accepted: date}
\maketitle

\begin{abstract}
  Germanium (Ge) detectors with ability of measuring a single electron-hole (e-h) pair are needed in searching for light dark matter (LDM) down to the MeV scale. We investigate the feasibility of Ge detectors with amorphous-Ge (a-Ge) contacts to achieve the sensitivity of measuring a single e-h pair, which requires extremely low leakage current. Three Ge detectors with a-Ge contacts are used to study the charge barrier height for blocking electrons and holes. Using the measured bulk leakage current and the D$\ddot{o}$hler-Brodsky model, we obtain the values for charge barrier height and the density of localized energy states near the Fermi energy level for the top and bottom contacts, respectively. We predict that the bulk leakage current is extremely small and can be neglected at helium temperature ($\sim$4 K). Thus, Ge detectors with a-Ge contacts possess the potential to measure a single e-h pair for detecting LDM particles.  

\end{abstract}

\section{Introduction}
\label{sec:intro}
 Light dark matter (LDM) especially low-mass dark matter in the MeV-scale has risen to become an exciting dark matter candidate in the past decade~\cite{ess2012, ess2016, ho, ste}. Despite various target materials having been used in the direct detection of Weakly Interacting Massive Particles (WIMPs, a popular dark matter candidate) for over three decades, all existing dark matter experiments are unable to detect MeV-scale dark matter since they are all sensitive to WIMPs with masses greater than a few GeV/c$^{2}$~\cite{cd09, cd, cd1, cd2, cog, cre12, cou, bar, dama, dar, dri, ede, kim, lux, pan, pic, cd14, xe11, xe15, xe17, xma, zep}. More recently, Kadribasic et al. have reported a method of using solid state detectors with directional sensitivity to dark matter interactions to detect low-mass WIMPs below 1 GeV/c$^{2}$ ~\cite{kad}. CRESST has achieved a threshold of 20 eV with a small prototype sapphire detector~\cite{cre17}. DAMIC has claimed a sensitivity to ionization $<$ 12 eV with silicon CCDs and consider their method to be able to reach 1.2 eV~\cite{dam17}.
 
 High-purity germanium (HPGe) detector technology has been used for dark matter searches since 1987 due to its high radio-purity~\cite{ahl}. Two main advantages of HPGe detector technology are its excellent energy resolution and high detection efficiency, which allow Ge detectors to reach a quite low energy threshold down to $\sim$0.5 keV$_{ee}$, where keV$_{ee}$ represents electronic equivalent energy. This enables Ge detectors to be used for detecting low mass WIMPs of a few GeV/c$^{2}$. More recently, a Ge detector utilizing internal charge amplification for the charge carriers created by the ionization of impurities has been demonstrated theoretically to be a promising new technology for detecting MeV-scale dark matter~\cite{mei}.
 
 A simple Ge detector is usually made of a block of HPGe crystal in which p$^{+}$ and n$^{+}$ electrical contacts have been fabricated on opposite sides of the block. The standard contact technology for Ge detectors are Li-diffused n$^{+}$ contact for hole blocking and B-implanted p$^{+}$ contact for electron blocking. This conventional contact technology is robust since both types of contacts are able to withstand high electric fields, and thus yield low charge carrier injection. However, there are drawbacks in this technology especially for the Li-diffused contacts. Due to its thickness and significant diffusion of Li at room temperature, the Li-diffused side is problematic for two main reasons. One is that the Li-diffused side usually forms a thick dead layer ($\sim$1 mm), which is insensitive to charge carriers, and a transition layer ($\sim$1 mm) between Li and Ge where charge carriers can be significantly trapped. The total thickness of the dead layer and the transition layer reduces the sensitive volume of a Ge detector. Another is that the Li-diffused side is difficult for segmentation in order to produce position-sensitive detectors, which are demanded increasingly by modern applications requiring not only excellent spectroscopy, but also particle tracking or imaging~\cite{looker}. 
 
\begin{figure} [htbp]
  \centering
  \includegraphics[width=0.45\linewidth]{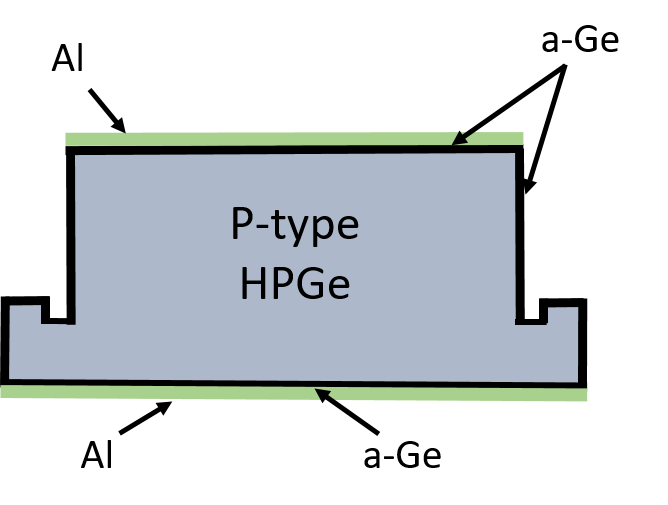}
  \includegraphics[width=0.3\linewidth]{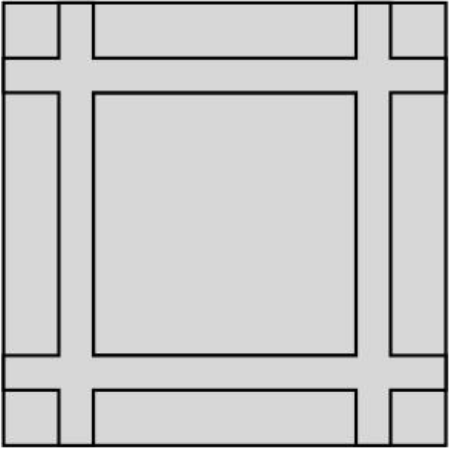}
  \caption{Schematic cross-sectional (left) and top view (right) drawing showing the typical geometry of the HPGe detectors without a guard-ring structure fabricated at USD.}
  \label{fig:geo}
\end{figure}

 An alternative contact technology capable of producing detectors with a thin contact without a dead layer and a transition layer, as well as with fine spacial resolution is the amorphous-germanium (a-Ge) contact developed at Lawrence Berkeley National Laboratory (LBNL)~\cite{luke3, amman1, amman2}. The successful use of a-Ge contacts on Ge detectors has been verified by a series of previous studies conducted at LBNL~\cite{luke1, luke2, luke4, hull, amman3, looker, amman4}. In addition, a large number of detectors with great detector performance have been fabricated successfully at LBNL using this technology. More recently, a dozen small detectors made from USD-grown crystals have also been fabricated successfully at the University of South Dakota (USD)~\cite{wei, meng} using the same technology. With a-Ge contact technology, as shown in Fig.~\ref{fig:geo}, a block of HPGe crystal was first coated with a thin film of high-resistivity a-Ge on all of its surfaces, then a thin film of low-resistivity metal (typically Al) was deposited on opposite sides of the crystal block. There are several advantages of this technology~\cite{amman5}: (1) simple fabrication process; (2) good bipolar blocking behavior, i.e. a-Ge contacts can block both hole and electron injection well; (3) thin contacts without a dead layer and a transition layer; (4) complete surface passivation, since a-Ge layers are common passivation materials; and (5) fine achievable contact pitches for segmentation. 
 
\begin{figure}
\includegraphics[width=0.48\textwidth]{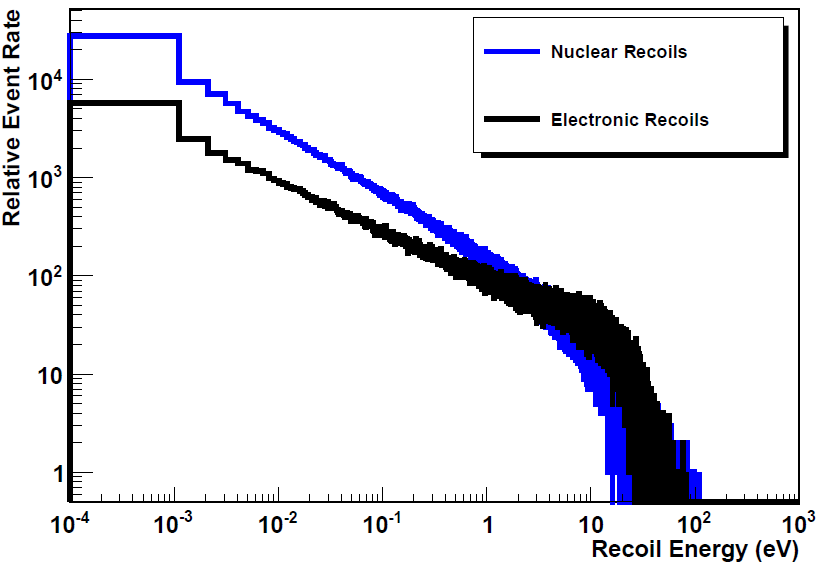}
\caption{
The relative event rate as a function of recoil energy for DM with masses 
between 0.1 MeV/c$^{2}$ to 1 GeV/c$^{2}$~\cite{mei}. 
}
\label{fig:fullSpe}
\end{figure}

 In order to directly detect MeV-scale dark matter, a detector with the ability to measure a single electron-hole (e-h) pair is required, since both electronic recoils and nuclear recoils induced by MeV-scale dark matter are in the range of sub-eV to 100 eV~\cite{ess2012}. Mei et al. has suggested direct detection of MeV-scale DM utilizing germanium internal charge amplification (GeICA) for the charge created by the ionization of impurities~\cite{mei}. GeICA can reach a detection energy threshold as low as 0.1 eV, allowing a large portion of both electronic recoils and nuclear recoils in the range of sub-eV to 100 eV~\cite{ess2012, mei} induced by DM to be accessible. For the DM-WIMPs interaction, the effective mass of WIMPs that contributed to the super-weak coupling strength is constrained between $\sim$1 MeV/c$^2$ to $\sim$100 MeV/c$^2$ by Xenon1T~\cite{dmwz}. Therefore, the detectable recoil energy is also in the range of sub-eV to 100 eV, similar to the recoil energy spectrum induced by MeV-scale DM as shown in Fig.~\ref{fig:fullSpe}~\cite{mei}. In this sub-eV to $\sim$100 eV region, GeICA offers a very competitive sensitivity to detect LDM.  

A conceptual GeICA detector can be made as a p-type point contact (PPC) detector. The size of the point contact can be as small as 0.5 mm to achieve low capacitance and high electric field. A low capacitance in sub picofarads (pF) is required to achieve ultra-low electronic noise. A high electric field near the point contact is needed to achieve the needed amplification factor for a GeICA detector. To achieve the sensitivity of detecting a single e-h pair induced by MeV-scale DM with GeICA detectors, one must be able to control the bulk leakage current much below 1 picoampere (pA). This is because one e-h pair corresponds to $\sim$pA if the drift time is on the level of $\sim$100 nanoseconds (ns) for a GeICA detector. For a GeICA detector with an amplification factor of 100, the size of signal would be 100 e-h pairs. This is equivalent to the number of e-h pairs produced by an electronic recoil with 300 eV energy in Ge assuming the average energy required to generate one e-h pair is 3 eV. Thus, low-noise electronics, as low as 40 eV-FWHM (full width at half maximum), demonstrated by Barton et al.~\cite{barton} would allow us to achieve the ability of measuring a single e-h pair with a GeICA detector. This indicates that the detector leakage current must be minimized since the leakage current mimics the signal and hence becomes a significant source of background.
 
\begin{figure} [htbp]
  \centering
  \includegraphics[width=0.4\linewidth]{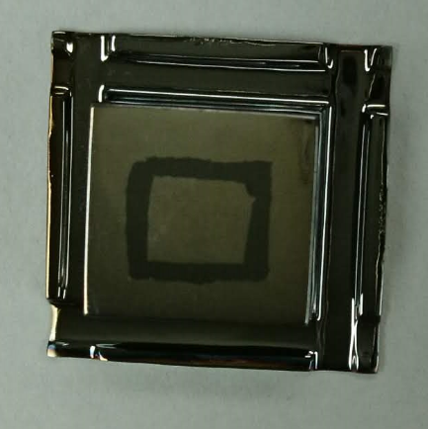}
  \includegraphics[width=0.5\linewidth]{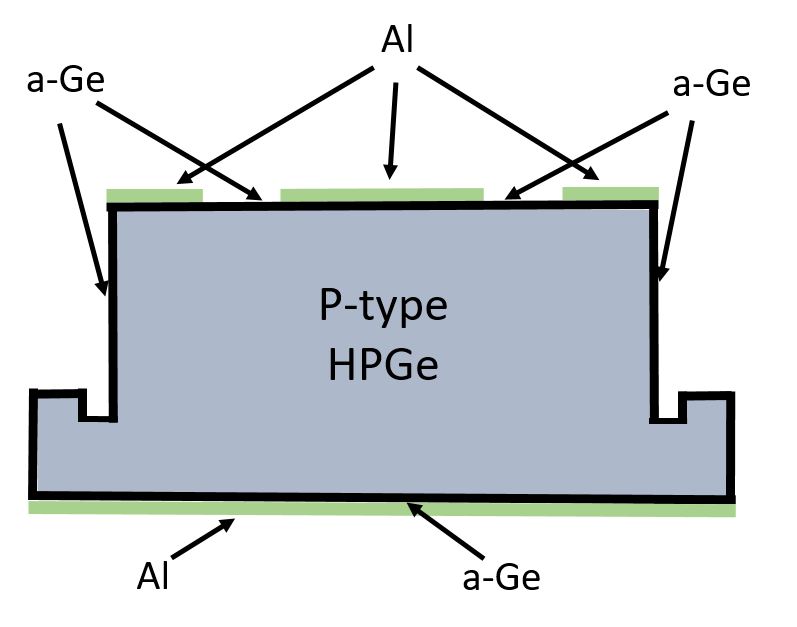}
  \caption{Left: The top view of a guard-ring detector, USD-R03, fabricated at USD; Right: The schematic cross-sectional view of a typical guard-ring detector fabricated at USD.}
  \label{fig:gud}
\end{figure}

 In general, there are three main contributions to the leakage current in a fully-depleted HPGe detector~\cite{looker}: (1) charge carrier injection at the electrical contact, i.e. hole injection at the positive contact and electron injection at the negative contact; (2) charge flow along detector side surfaces; and (3) thermal generation of electron-hole pairs in the detector. The contribution from the thermal generation can be reduced to a negligible level by cooling the detector to liquid helium temperature. According to the studies in ~\cite{hull,looker}, a guard ring structure shown in Fig.~\ref{fig:gud} can be used to mitigate the contribution to the measurement from the surface current component, leaving charge carrier injection at the electrical contacts as the main source of leakage current. It is worth mentioning that Barton et al. has demonstrated a low-leakage current of 0.02 pA at 30 K for a PPC detector with bias of 150 V~\cite{barton}. The outer surface of this PPC detector was made with lithium diffused n-type contact for hole blocking and the inner point contact was made with the bipolar blocking of amorphous silicon. For a Ge detector made with a-Ge contact technology, the level of charge injected into the detector is dictated by the electron or hole energy barrier to charge injection, which is an important property of a-Ge contacts. The barrier heights of the a-Ge contacts depend on the fabrication method used to produce them. Therefore, a study of the electron or hole energy barrier to charge injection is necessary to optimize fabrication parameters and thus improve the detector performance especially in terms of reducing the leakage current, so that the detector will be able to reach low-energy threshold for MeV-scale dark matter detection.
 
 In this paper, we report the electron and hole energy barrier heights of RF-sputtered a-Ge contacts on three HPGe detectors with guard-ring structures fabricated at USD. All three detectors were made with the Ge crystals grown by us at USD. The thickness of the a-Ge layer on the tom/bottom and the side surfaces are measured using the Alpha-Step Profiler (KLA Tencor) in our lab. The method was described in detail in an earlier publication~\cite{meng}. For a 30-min deposition time used for the three detectors reported in this work, the thickness of the a-Ge layer on the top/bottom and the side surfaces are around 1.2 um and 556 nm, respectively. Note that the theoretical model used in this work is independent of the thickness of the a-Ge layer. A theoretical model of the amorphous semiconductor electrical contact is presented in section~\ref{sec:theo}, followed by the experimental methods in section~\ref{sec:exp}. The data analysis and results are discussed in section~\ref{sec:dis}, followed by the discussion and prediction in section~\ref{sec:pred}. Finally, the conclusions are summarized in section~\ref{sec:conl}.
 
 \section{Theoretical Model of Amorphous Semiconductor Electrical Contract}
\label{sec:theo}
To implement strategies to optimize the a-Ge contact and thus minimize the detector leakage current, it is of great importance to understand the underlying physics of the amorphous semiconductor contacts. Thus, a main focus in this paper is to analyze our detector leakage current data based on a physical model of the amorphous semiconductor electrical contact.

Compared to a crystalline semiconductor, an amorphous semiconductor is a solid that lacks the long-range crystalline order of atoms. Amorphous semiconductor films are usually formed through evaporation~\cite{han}, sputtering~\cite{luke1}, or damage to a crystal lattice~\cite{pru}, while single-crystal semiconductors are produced using the Czochralski technique~\cite{mot1} or other crystal growing techniques, which require much more time and effort. Although atoms in amorphous semiconductors still arrange in a diamond-like structure, the density of defects states in amorphous semiconductors is much higher than that of crystalline semiconductors because of inefficient stacking of those diamond-like structures. Many of those defects are vacancies or voids, leaving "dangling bonds", which is a semiconductor valence state not occupied in a covalent bond~\cite{mot2}. Since the binding energy of dangling bonds is less than that of covalent bonds, the dangling bonds more easily contribute to electrical conduction. There are large number of defect states in the band gap of amorphous semiconductors~\cite{mot2}. Because of these defect states, charge carriers that would normally be forbidden to move in the gap region can move from defect to defect in a process called phonon-assisted hopping, which allows significant electrical conduction near the Fermi level in the gap region of amorphous semiconductors. This hopping conduction through the localized defect energy states near the Fermi level is the dominant source of charge movement in amorphous semiconductors~\cite{brod2}. 

In fact, the charge conduction in a metal also mainly occurs through electronic energy levels close to the Fermi energy. This similarity in conduction motivates the use of metal-semiconductor theory or Schottky theory~\cite{sze} to describe the amorphous semiconductor contacts. A theoretical model about the current-voltage relationship for amorphous-crystalline semiconductor heterojunctions has been developed by D$\ddot{o}$hler and Brodsky during the 1970s~\cite{brod2, dohl,brod1}. D$\ddot{o}$hler and Brodsky concluded that the forward biased junction should be indistinguishable from those for a metal semiconductor junction and the reverse current should have no saturation but should show an exponential increase as the space charge lowers the barrier height. In the D$\ddot{o}$hler-Brodsky model, the state of thermal equilibrium in which the Fermi levels in the amorphous semiconductor and crystalline semiconductor coincide and an ideal contact without interface energy states was assumed. Thus, as described clearly in the energy diagrams presented in the most recent work by Amman~\cite{amman4}, for the charge carriers at the Fermi level to be injected from the amorphous semiconductor to the conduction band of the crystalline semiconductor, there exists a potential energy barrier that inhibits such injection. This is the physical mechanism that leads to the bipolar blocking behavior of amorphous semiconductor contacts. The D$\ddot{o}$hler-Brodsky model was later applied successfully to the a-Ge contact on HPGe in the studies by Hull et al.~\cite{hull}, Looker et al.~\cite{looker} and Amman~\cite{amman4}. Based on the D$\ddot{o}$hler-Brodsky model and Schottky theory~\cite{sze,heni}, for a p-type HPGe detector fabricated with a-Ge contacts when a negative bias voltage is applied to one of its contacts, the current normalized by the contact area, $J$, is given by~\cite{looker, amman4}:
\begin{multline}
J = A^*T^2exp[-(\varphi_h-\Delta\varphi_h)/kT] 
\\ with \quad \Delta\varphi_h = \sqrt{2qV_aN_d/N_f}
  \label{e:j1}    
\end{multline}

and,

\begin{multline}
J = A^*T^2exp[-(\varphi_e-\Delta\varphi_e)/kT]
\\ with \quad \Delta\varphi_e = \sqrt{\varepsilon_0\varepsilon_{Ge}/N_f}(V_a-V_d)/t
  \label{e:j2}    
\end{multline}
where Eq.~\ref{e:j1} describes the leakage current density from hole injection at the contact where the detector starts to deplete, and Eq.~\ref{e:j2} describes the leakage current density, after the detector is fully depleted, from electron injection at another contact where a negative bias voltage is applied. The overall leakage current above full depletion is then the sum of Eq.~\ref{e:j1} and Eq.~\ref{e:j2}. In the equations above, $A^*$  is the effective Richardson constant, $T$ is the temperature, $\varphi_h$ and $\varphi_e$ are the energy barriers to hole and electron injections, respectively, $\Delta\varphi_h$ and $\Delta\varphi_e$ are the barrier lowering terms which account for the lowering of hole and electron energy barrier, respectively, due to the penetration of the electric field into the a-Ge contacts, $q$ is the magnitude of the electron charge, $V_a$ is the reverse bias voltage applied across the detector, $N_d$ is the net ionized impurity concentration of the detector, $N_f$ is the density of localized energy states (defects) near the Fermi level in the a-Ge, $k$ is the Boltzmann constant, $\varepsilon_0$ is the free-space permittivity, $\varepsilon_{Ge}$ is the relative permittivity for Ge, $V_d$ is the full depletion voltage and $t$ is the detector thickness. Note that both Eq.~\ref{e:j1} and Eq.~\ref{e:j2} above have been adapted from the expressions of Looker et al.~\cite{looker} and Amman~\cite{amman4} to our data analysis in this work. The only difference is that the original factor of $A^*$ predicted by the Schottky theory is calculated using the theory from S. M. Sze~\cite{sze} where 
\begin{equation}
A^{*} = \frac{4\pi q m^*k^2}{h^3}
\end{equation}
with $q$ being the electric charge, $m^{*}$ the effective mass of charge carriers, $k$ the Boltzmann constant and $h$ the Plank constant. Note that $A^{*}$ is the effective Richardson constant for thermionic emission, neglecting the effects of optical-phonon scattering and quantum mechanical reflection. For free electrons ($m^{*}$ = $m_{0}$), the Richardson constant $A$ is 120 Amp/cm$^2$/K$^2$~\cite{sze}. For p-type Ge in the <100>-direction~\cite{sze}, 
\begin{equation}
\frac{A^{*}}{A}=\frac{m^{*}_{lh}+ m^{*}_{hh} +m^{*}_{h,so}}{m_{0}},
\end{equation} 
where $m^{*}_{lh} = 0.04 m_{0}$, $m^{*}_{hh} = 0.28 m_{0}$, and $m^{*}_{h,so} = 0.08 m_{0}$~\cite{zeg}. Thus, $A^{*}$ = 48 Amp/cm$^2$/K$^2$, which has been used in~\cite{hhl}. As discussed by pioneers in early publications~\cite{hull, looker, amman4}, this effective Richardson constant may vary with the fabrication processes. However, we show that $A^{*}$ = 48 Amp/cm$^2$/K$^2$ in this work to explain the behavior of our detector leakage current especially from the hole injection at the contact when the detector is partially depleted. 

Eq.~\ref{e:j1} and Eq.~\ref{e:j2} clearly show that, at a given temperature, the energy barrier height plays a key role in the effectiveness of the charge injection blocking behavior of the amorphous electrical contacts. Based on the metal-semiconductor theory, the energy barrier height is, in general, determined by the work function of amorphous semiconductors, the electron affinity of crystalline semiconductors, and the interface states between amorphous and crystalline semiconductors~\cite{sze}. Since the magnitude and energy distribution of the interface states are likely dependent on the preparation of the crystalline semiconductor surface and the deposition process of the amorphous semiconductor, it is not straightforward to determine the energy barrier height theoretically. In the data analysis of this work, we followed the method adapted by Amman~\cite{amman4} and treated the barrier height as a parameter to be determined from electrical measurements. To measure the barrier height from the data, Eqs.~\ref{e:j1} and ~\ref{e:j2} can be rewritten as:
\begin{equation}
kTln(\frac{J}{A^*T^2})=-\varphi_h+b_1\sqrt{V_a} \quad \textrm{with} \quad b_1=\sqrt{2qN_d/N_f}
  \label{e:j3}
\end{equation}

and,

\begin{multline}
kTln(\frac{J}{A^*T^2})=-\varphi_e+b_2(V_a-V_d)
\\ with \quad b_2=\sqrt{\varepsilon_0\varepsilon_{Ge}/N_f}/t
  \label{e:j4}    
\end{multline}

Eq.~\ref{e:j3} indicates that, with constants of $k$, $A^*$, $q$ and $N_d$, $\varphi_h$ and $N_f$ for the contact where the detector starts to deplete can be estimated by fitting the measurements of $J$-$V_a$ below full depletion at a given temperature, $T$. Using the parameters extracted from the fit to Eq.~\ref{e:j3}, the hole contribution to leakage current can be estimated for all applied voltage values and subtracted from the $J$-$V_a$ data. With another fit to Eq.~\ref{e:j4}, $\varphi_e$ and $N_f$ for the other contact can be also determined.

\section{Experimental Methods}
\label{sec:exp}

\subsection{Detector fabrication}
For the study in this work, we have fabricated three planar Ge detectors, USD-R02, USD-R03 and USD-W03, with a guard-ring structure on the top surface of each detector. As an example, Fig.~\ref{fig:gud} (left) shows the top view of one of the guard-ring detectors, USD-R03. Fig.~\ref{fig:gud} (right) presents the schematic cross-sectional view of a typical guard-ring detector fabricated at USD. For the details about how to convert an HPGe crystal into a planar detector with a guard-ring structure shown in Fig.~\ref{fig:gud}, please refer to our previous work~\cite{meng, wei}. Due to more uncertain handling processes involved before the a-Ge coating on the bottom contact, it is expected that the top and bottom contacts may not have the same quality. The dimensions of the three guard-ring detectors fabricated at USD for this study are shown in Table~\ref{tab:dim}. 
\begin{table*}[t]
  \centering
  \caption{The dimensions of the three guard-ring detectors, USD-R02, USD-R03 and USD-W03, fabricated at USD in this work.}
 \begin{tabular}{p{1.7cm}p{1.5cm}p{1cm}p{1cm}p{1cm}p{1cm}p{2.5cm}}
\hline
\multicolumn{1}{c}{\multirow{2}{1cm}{}} & \multirow{2}{*}{} &  \multicolumn{2}{c}{Bottom} & \multicolumn{3}{c}{Top}  \\ \cline{3-7} 
\multicolumn{1}{c}{Detectors} &Thickness (cm) & Length (cm)& Width (cm) & Outer length (cm) & Outer width (cm) & Center contact area (cm$^2$)\\ \hline
 USD-R02& 0.65  & 2.3 & 2.2 & 1.6 & 1.4 & 0.29   \\ \hline
 USD-R03& 0.81  & 2.3 & 2.3 & 1.6 & 1.6 & 0.48  \\ \hline
 USD-W03& 0.94  & 2.0 & 1.92 & 1.21 & 1.16 & 0.24  \\ \hline
  \end{tabular}
  \label{tab:dim}
\end{table*}

\subsection{Detector characterization}
\begin{figure} [htbp]
  \centering
  \includegraphics[clip,width=0.6\linewidth]{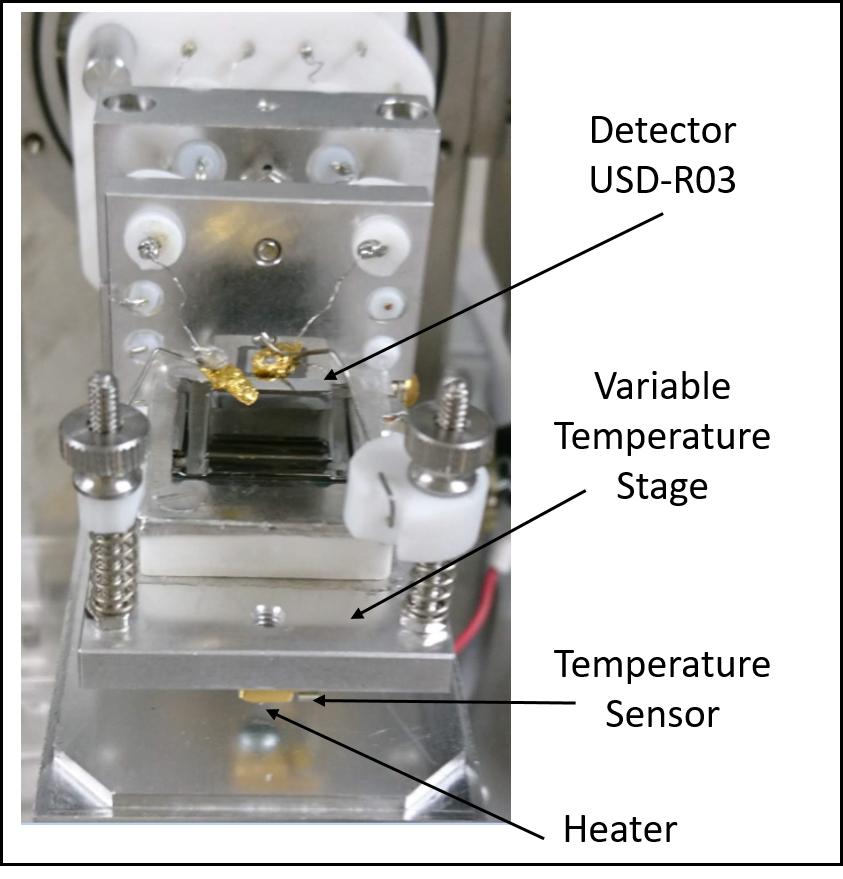}
  \caption{Shown is the detector USD-R03 loaded in a variable-temperature cryostat.}
  \label{fig:cryo}
\end{figure}

\begin{figure} [htbp]
  \centering
  \includegraphics[clip,width=0.9\linewidth]{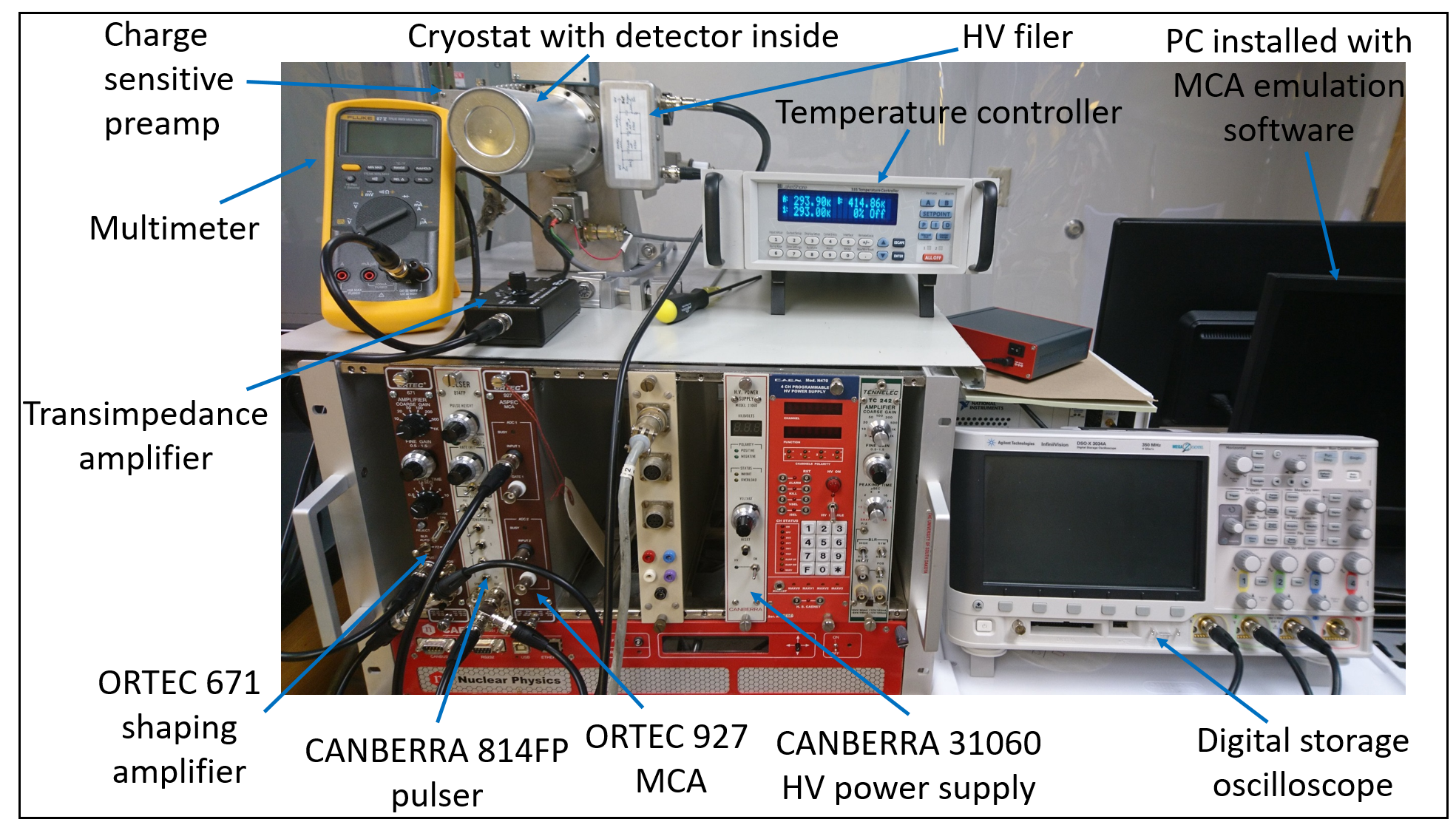}
  \caption{Experimental setup for detector characterization.}
  \label{fig:setup}
\end{figure}

\begin{figure} [htbp]
  \centering
  \includegraphics[clip,width=0.9\linewidth]{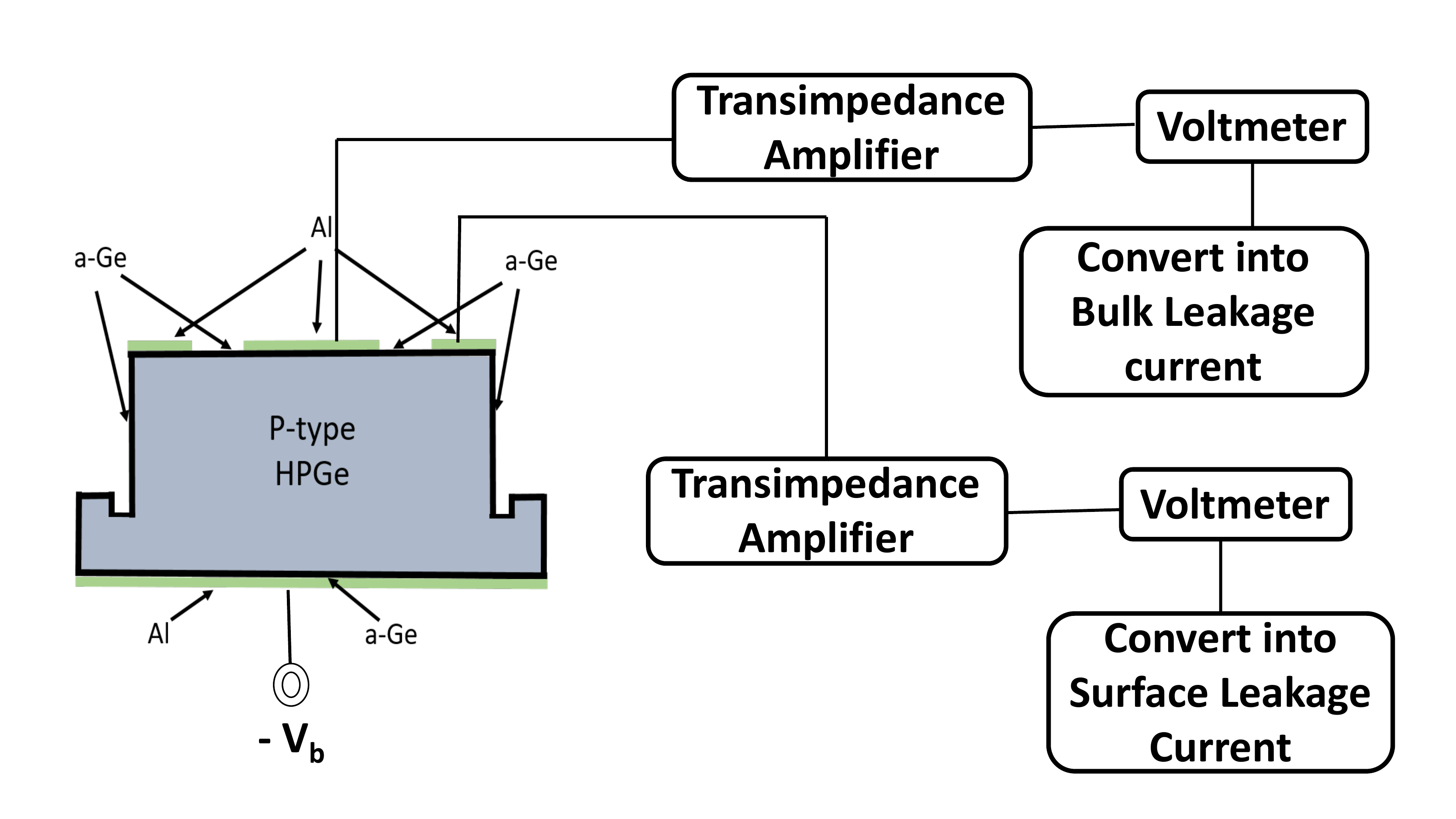}
  \caption{Shown is a sketch of the setup for measuring the bulk leakage current and the surface leakage current using the electronics depicted in Fig.~\ref{fig:setup}. The detector was virtually grounded through the transimpedance amplifier, which allows us to measure the current down to pA. }
  \label{fig:dia}
\end{figure}

After each detector was fabricated, it was loaded in a variable-temperature sample cryostat as shown in Fig.~\ref{fig:cryo}. This cryostat is provided by LBNL. The small size of this cryostat makes for faster vacuum pumping and cooling to base temperature. The variable-temperature stage shown in Fig.~\ref{fig:cryo} has a temperature sensor and a small heater attached so that the temperature of this stage can be uniformly elevated above the liquid nitrogen temperature. The base temperature is about 79 K, while the maximum is in excess of 200 K~\cite{looker} with the LakeShore 335 temperature controller shown in Fig.~\ref{fig:setup}. Also shown in Fig.~\ref{fig:setup} are the required signal processing electronics and measurement electronics to conduct electrical and spectroscopy measurements. The diagram presented in Fig.~\ref{fig:dia} shows how we performed the electrical and spectroscopy measurements using the electronics shown in Fig.~\ref{fig:setup}. The detector was virtually grounded through the transimpedance amplifier. 

\begin{figure} [htbp]
  \centering
  \includegraphics[clip,width=0.9\linewidth]{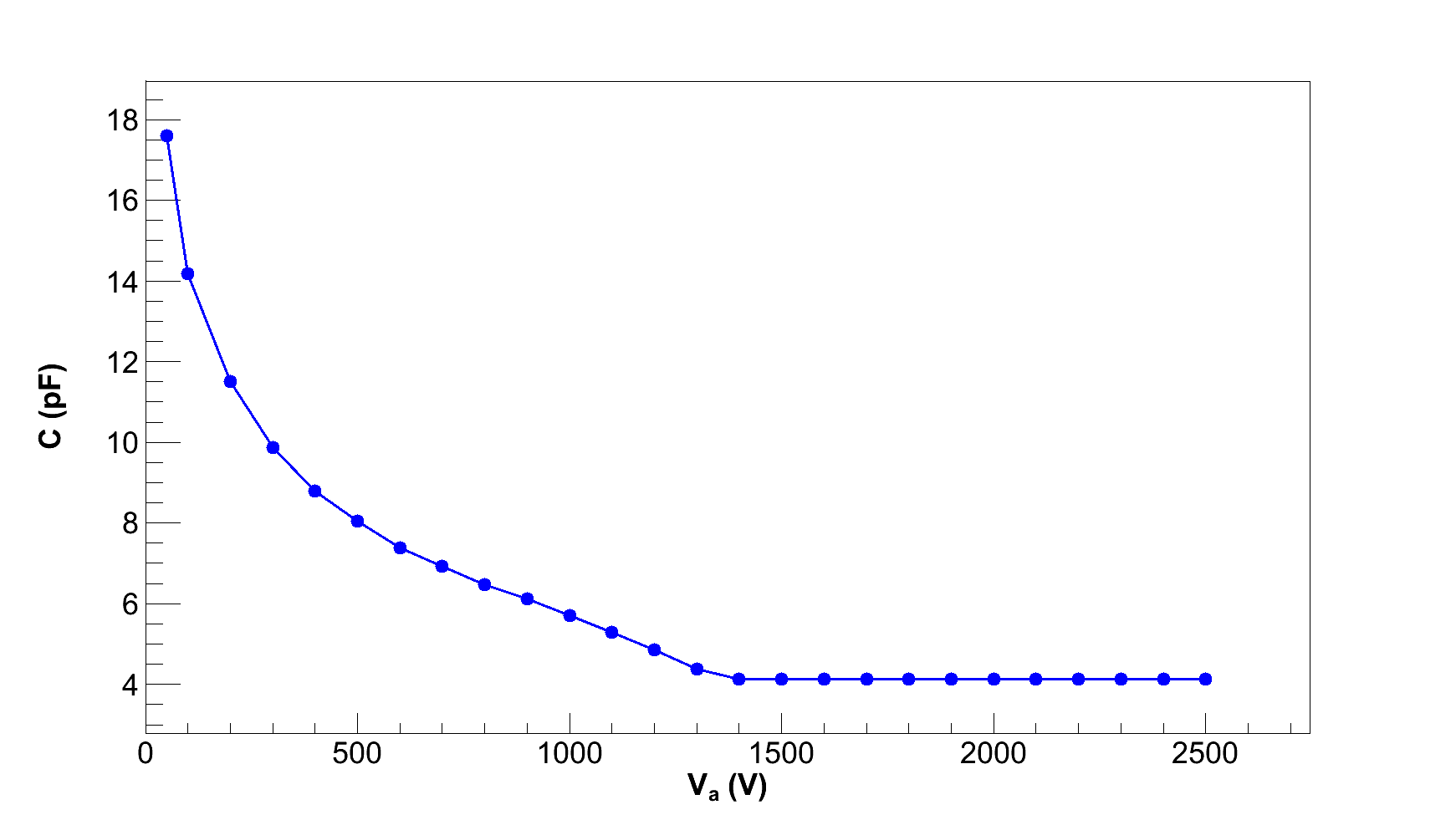}
  \caption{Measured detector capacitance as a function of bias voltage for detector USD-R03.}
  \label{fig:cv}
\end{figure}

\begin{figure} [htbp]
  \centering
  \includegraphics[clip,width=0.9\linewidth]{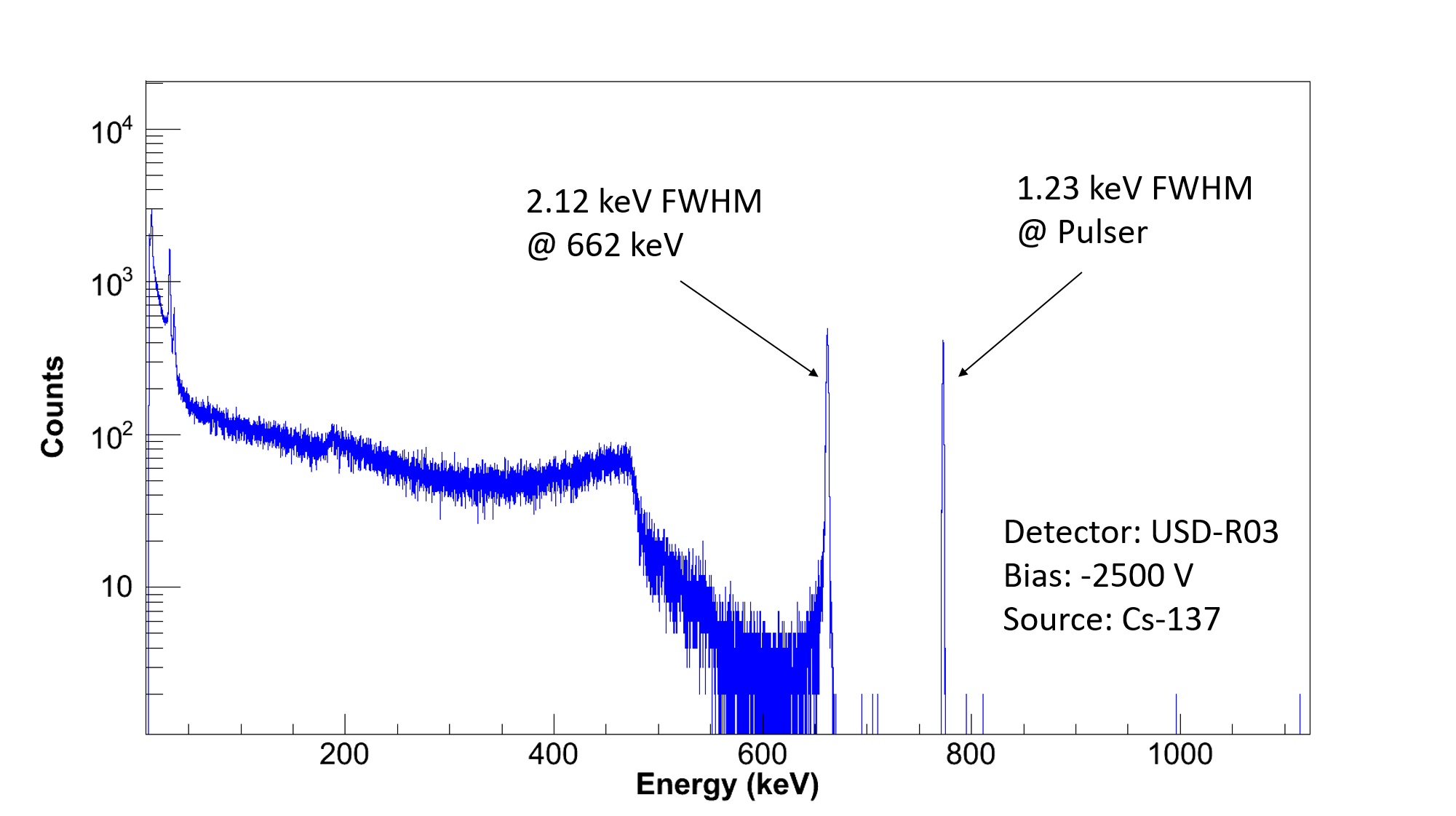}
  \caption{Energy spectrum from a Cs-137 source measured with the detector USD-R03. The source was positioned facing the detector bottom. The bias voltage of -2500 V was applied to the bottom electrical contact on the detector while the signals were measured from the top.}
  \label{fig:cs137}
\end{figure}

To extract the barrier height by using the theoretical model (Eqs.~\ref{e:j3} and ~\ref{e:j4}) described in section~\ref{sec:theo}, the $I$-$V$ (current-voltage) curve with an indication of the full depletion voltage needs to be measured for each detector, so that the hole and electron barrier heights can be studied separately. According to our previous study~\cite{wei}, such an $I$-$V$ curve can be obtained when the detector operating temperature is higher than 79 K, such as 90 K or higher. To validate the current-voltage data from each detector used in this work, it is necessary to verify that each detector is a workable detector. The following measurements have been conducted for each detector operated at $\sim$79 K for this verification: (1) leakage current as a function of the bias voltage ($I$-$V_a$) from both guard-ring and center contacts; (2) capacitance as a function of the bias voltage ($C$-$V_a$) from center contact; (3) spectroscopy measurement with a radiation source of Cs-137 from center contact. The leakage current level reflects the quality of the a-Ge contacts. The $C$-$V_a$ measurements allowed us to determine the full depletion voltage of the detector and the impurity concentration of the crystal. The energy resolution information can be obtained from the spectroscopy measurement. More details about how we conducted the electrical ($I$-$V_a$, $C$-$V_a$) and spectroscopy measurements with the electronics shown in Fig.~\ref{fig:setup} are stated in our previous work ~\cite{wei}. The testing results including the full depletion voltage ($V_d$), the leakage current at full depletion voltage ($I_d$) at the center contact, crystal impurity concentration from $C$-$V_a$ measurements, the full width at half maximum (FWHM) at 662 keV and the FWHM of a pulser peak, which determines the noise level of the detector, are shown in Table~\ref{t:detectors} for each detector. As an example, Fig.~\ref{fig:cv} and Fig.~\ref{fig:cs137} show the $C$-$V_a$ curve and the energy spectrum of Cs-137 source measured by the detector USD-R03 at 79 K. Also shown in Fig.~\ref{fig:cs137} is an artificial peak due to the injected pulses from the high voltage line to measure the detector noise in terms of FWHM.
\begin{table*}[t]
  \centering
  \caption{A summary of detector performance for three guard-ring detectors used in this work. $V_d$ and $I_d$ denote the full depletion voltage and the measured center contact leakage current at full depletion voltage, respectively. The FWHM at the pulser peak represents the detector noise level.}
 \begin{tabular}{p{1.7cm}p{1.2cm}p{1.3cm}p{3.75cm}p{1.75cm}p{2.1cm}} \hline
    Detectors& $V_d$ (V) &$I_d$ (pA) &Crystal impurity concentration from $C$-$V$ measurements (/cm$^3$)&FWHM at 662 keV (keV)&FWHM of pulser peak (keV)  \\ \hline 
    USD-R02 &700 &1 &2.93$\times$10$^{10}$ &1.57 &1.01  \\ \hline
    USD-R03 &1400 &1 &3.78$\times$10$^{10}$ &2.12 &1.23  \\ \hline
    USD-W03 &1300 &1 &2.60$\times$10$^{10}$ &2.35 &1.33  \\ \hline
  \end{tabular}
  \label{t:detectors}
\end{table*}

\section{Data Analysis and Results}
\label{sec:dis}
When collecting leakage-current data at a given temperature for each detector, the detector depletion process can either start from the top or bottom contact depending on the polarity of the bias voltage applied to the detector. With the detector configuration shown in Fig. ~\ref{fig:cryo}, a p-type detector starts to deplete from the top (bottom) contact with a negative (positive) bias voltage applied to the bottom contact. As mentioned in section~\ref{sec:theo}, if a detector can be fully depleted with the depletion starting from the top contact, one can determine $\varphi_h$ and $N_f$ for the top contact and $\varphi_e$ and $N_f$ for the bottom contact by fitting the $J$-$V_a$ data to Eqs.~\ref{e:j3} and ~\ref{e:j4}. Similarly, if a detector can be fully depleted with the depletion starting from the bottom contact, then $\varphi_h$ and $N_f$ for the bottom contact and $\varphi_e$ and $N_f$ for the top contact can be determined as well. That is, $\varepsilon_h$, $\varepsilon_e$ and $N_f$ can all be estimated if the detector can be fully depleted when depleting from both top and bottom contacts.  

\begin{figure} [htbp]
  \centering
  \includegraphics[clip,width=0.9\linewidth]{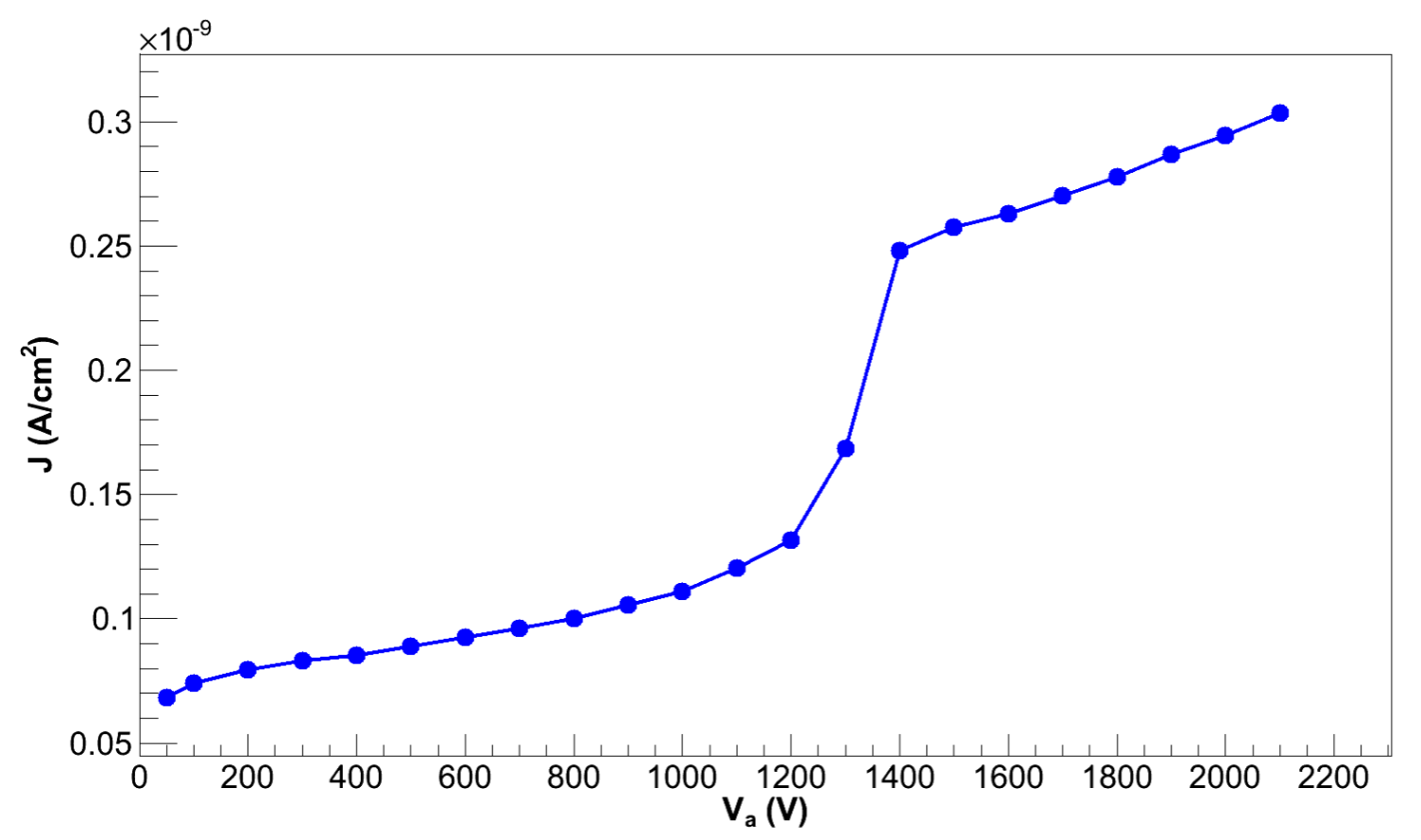}
  \caption{Measured leakage current density plotted as a function of bias voltage at 90 K for detector USD-R03 with the configuration shown in Fig.~\ref{fig:cryo}.}
  \label{fig:jv}
\end{figure}

Based on our $J$-$V_a$ measurements, detectors USD-R02 and USD-R03 can be fully depleted only when the depletion started from the top contact since the bottom contact cannot withstand high electric field penetration. For detector USD-W03, it can be fully depleted with depletion starting from both top and bottom contacts. Thus, for both detectors USD-R02 and USD-R03, we are able to determine $\varphi_h$ and $N_f$ for the top contact and $\varphi_e$ and $N_f$ for the bottom contact, while for detector USD-W03, we are able to determine $\varepsilon_h$, $\varepsilon_e$ and $N_f$ for each contact. In this work, the detector USD-R03 was used as an example to show how $\varphi_h$ and $N_f$ for the top contact and $\varphi_e$ and $N_f$ for the bottom contact were determined. Fig.~\ref{fig:jv} shows the measured center contact leakage current density as a function of bias voltage at 90 K from the detector USD-R03. 

\begin{figure} [htbp]
  \centering
  \includegraphics[clip,width=0.9\linewidth]{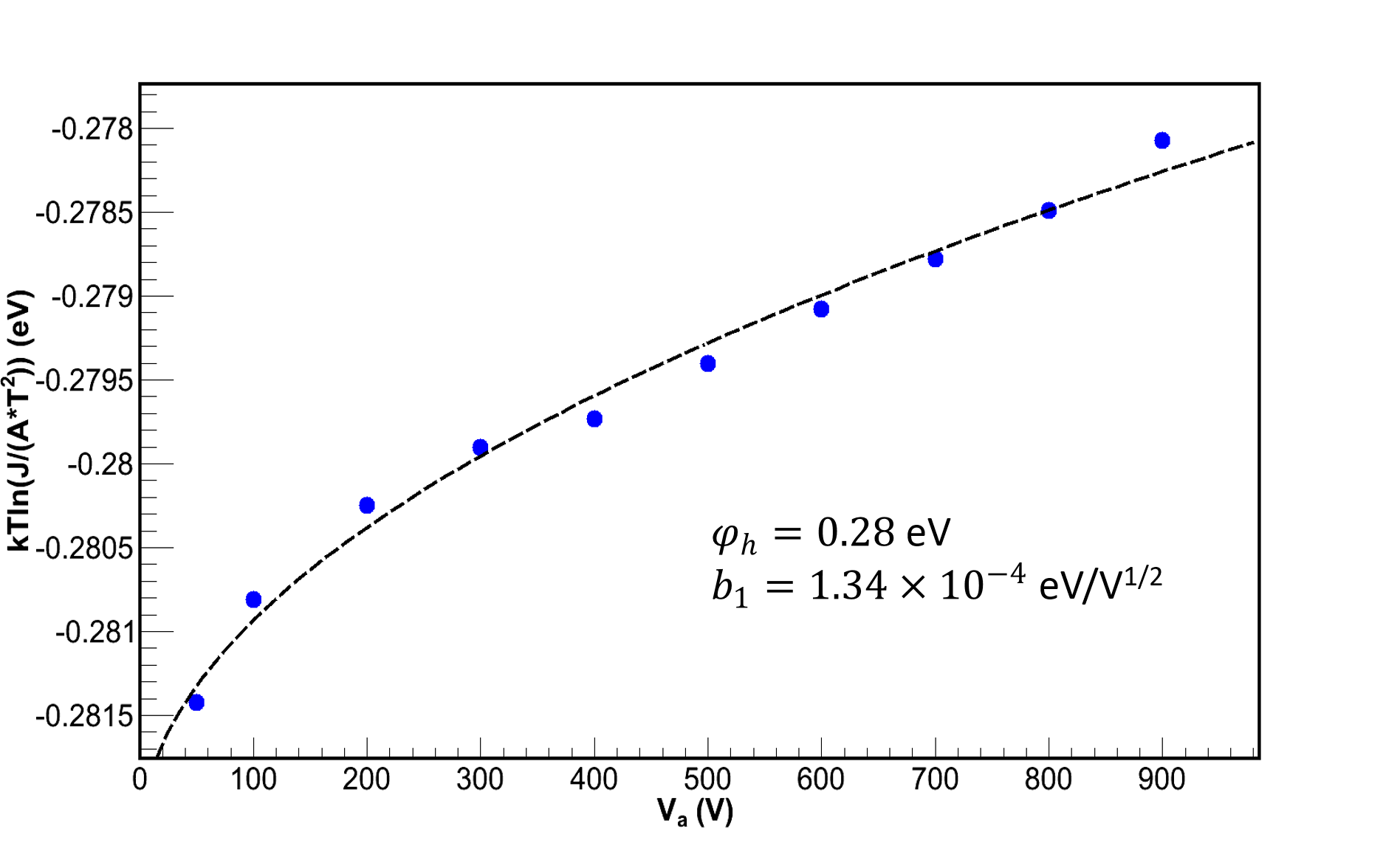}
  \caption{A fit of the plotted data of $kTln(J/(A^*T^2))$ as a function of $V_a$ below full depletion to Eq.~\ref{e:j3}. Based on the fit, the two free parameters in Eq.~\ref{e:j3} were determined to be, $\varphi_h$ = 0.28 eV and $b_1$ = 1.34$\times$10$^{-4}$/eV/V$^{1/2}$.}
  \label{fig:h90}
\end{figure}

\begin{figure} [htbp]
  \centering
  \includegraphics[clip,width=0.9\linewidth]{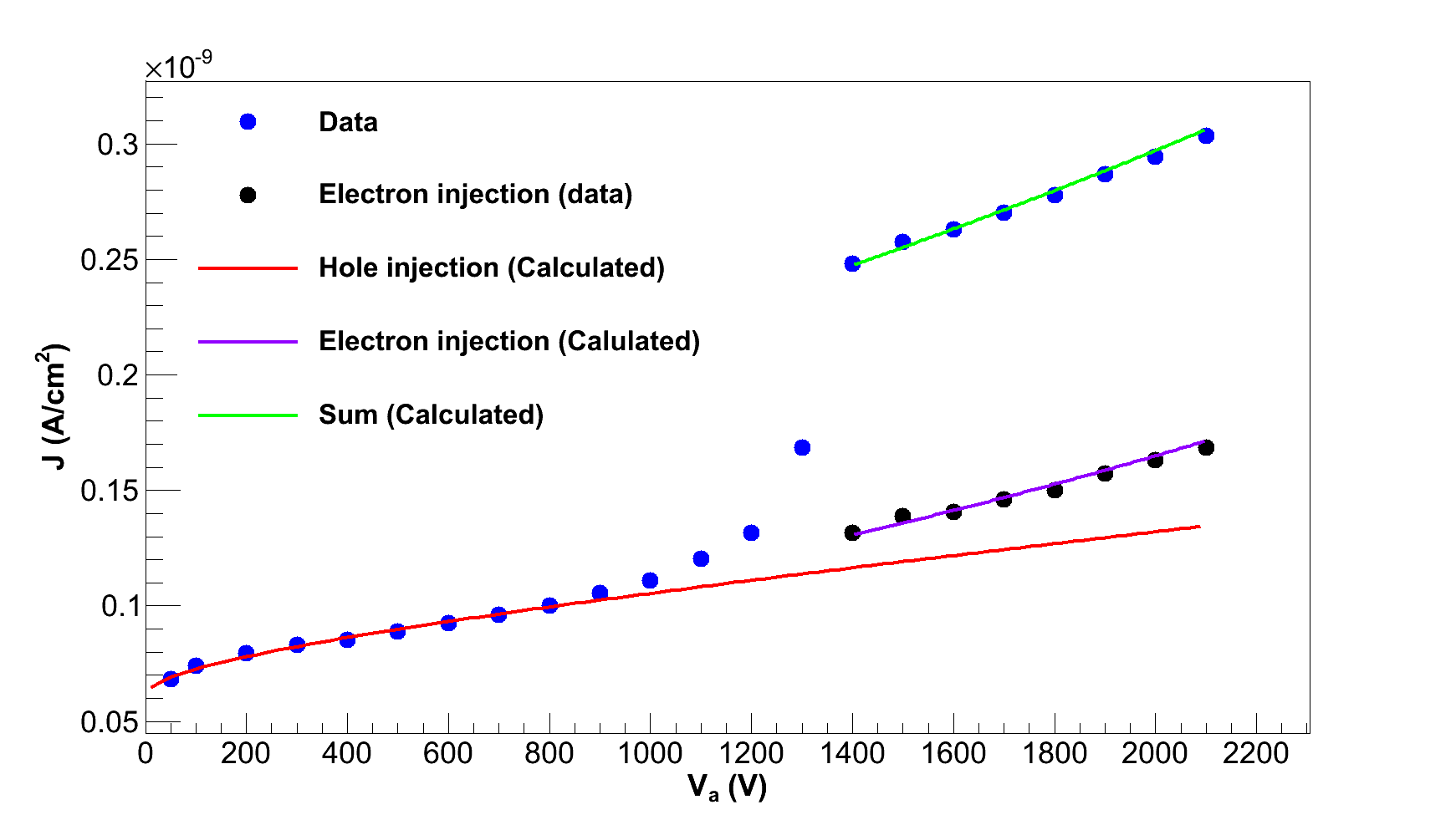}
  \caption{A comparison between the data and the theoretical models described in Eqs.~\ref{e:j1} and ~\ref{e:j2}. The blue dots are the measured $J$-$V$ data shown in Fig.~\ref{fig:jv}. The red line is the hole contribution to the leakage current calculated by Eq.~\ref{e:j3} with the fit parameters provided by Fig.~\ref{fig:h90}. The black dots are the leakage current purely from electron contribution by subtracting the hole contribution from the total leakage current. The purple line is the electron contribution to the leakage current calculated by Eq.~\ref{e:j4} with the fit parameters provided by Fig.~\ref{fig:e90}. The green line is the sum of electron and hole contributions calculated by Eq.~\ref{e:j3} and Eq.~\ref{e:j4}.}
  \label{fig:comp}
\end{figure}

\begin{figure} [htbp]
  \centering
  \includegraphics[clip,width=0.9\linewidth]{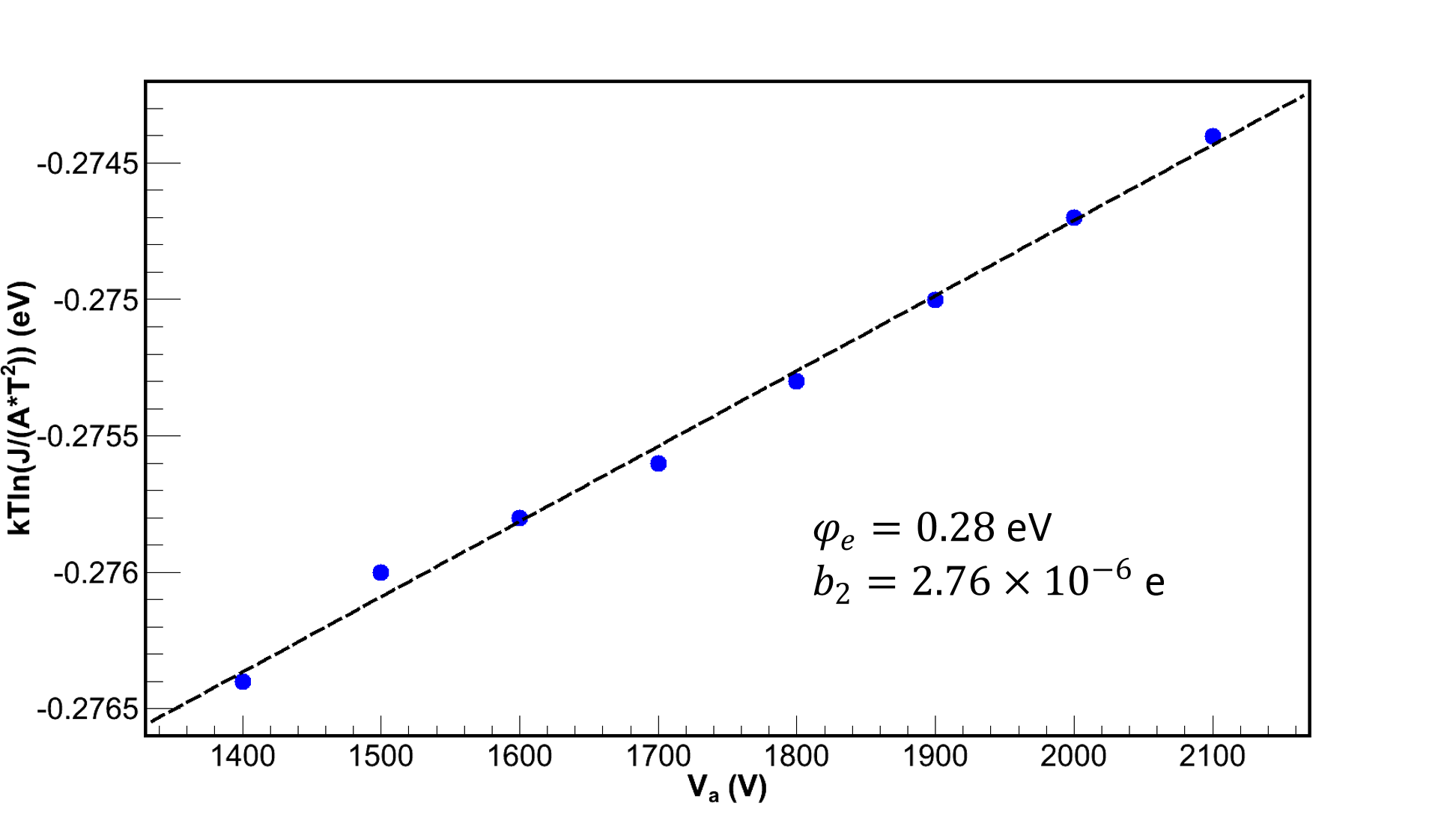}
  \caption{A fit of the plotted data of $kTln(J/(A^*T^2))$ as a function of $V_a$ above full depletion to Eq.~\ref{e:j4}. Based on the fit, the two free parameters in Eq.~\ref{e:j4} were determined to be, $\varphi_e$ = 0.28 eV and $b_2$ = 2.76$\times$10$^{-6}$ e.}
  \label{fig:e90}
\end{figure}

By fitting the plotted data of $kTln(J/(A^*T^2))$ as a function of $V_a$ below full depletion to Eq.~\ref{e:j3} with $k$ = 8.62$\times$10$^{-5}$ eV/K, $T$ = 90 K, $A^*$ = 48 A/cm$^2$/K$^2$, $q$ = 1.6$\times$10$^{-19}$ C, $N_d$ = 3.78$\times$10$^{10}$ cm$^{-3}$, as shown in Fig.~\ref{fig:h90}, we were able to determine $\varphi_h$ and $N_f$ to be, $\varphi_h$ = 0.28 eV and $N_f$ = 4.23$\times$10$^{18}$ eV$^{-1}$cm$^{-3}$, for the top contact. Using the parameters extracted from the fit shown in Fig.~\ref{fig:h90}, the hole contribution to leakage current can be estimated for all applied voltage values (red line in Fig.~\ref{fig:comp}) and subtracted from the $J$-$V_a$ data. At this point, only leakage current from electron injection at the bottom contact remains and there should be no current below full depletion voltage, which are the black dots in Fig.~\ref{fig:comp}. The values of $\varphi_e$ = 0.28 eV and $N_f$ = 1.50 $\times$10$^{18}$ eV$^{-1}$cm$^{-3}$ for the bottom contact can then be obtained by fitting the plotted data of $kTln(J/(A^*T^2))$ as a function of $V_a$ after full depletion to Eq.~\ref{e:j4} with $V_d$ = 1400 V, $\varepsilon_0$ = 8.85$\times$10$^{-14}$ C/V/cm and $\varepsilon_{Ge}$ = 16 and $t$ = 0.81 cm, as shown in Fig.~\ref{fig:e90}. To reduce the systematic uncertainty in the determination of $\varphi_h$, $\varphi_e$ and $N_f$, $J$-$V_a$ characteristics were measured at several different temperatures for each detector. Table~\ref{t:results} shows the average values of $\varphi_h$, $\varphi_e$ and $N_f$ for all temperatures for the corresponding a-Ge contact on each detector. The error quoted for each value represents the difference between the average value and the individual value obtained at different temperatures. The small differences among the values of $\varphi_h$, $\varphi_e$ and $N_f$ for each of the guard-ring detector indicates that our detector fabrication process is consistent and reliable (see Table~\ref{t:results}). 

\begin{table*}[t]
  \centering
  \caption{The values of $\varphi_h$, $\varphi_e$ and $N_f$ determined by fitting the plotted data of $kTln(J/(A^*T^2))$ as a function of $V_a$ to eqs.~\ref{e:j3} and ~\ref{e:j4} for the corresponding a-Ge contact on each detector. For the top contact of the detector, USD-R02, due to a flat distribution of the $J$-$V_a$ data before full depletion, it was safe to assume no electrical field penetration into the contact, i.e. $\Delta\varphi_h$ = 0 or $N_f \rightarrow \infty$.}
  \begin{tabular}{p{1.7cm}p{5cm}p{5cm}} \hline
   Detectors& Top Contact & Bottom Contact  \\ \hline 
    USD-R02 & $\varphi_h$ = (0.31$\pm$0.01) eV & $\varphi_e$ = (0.30$\pm$0.00) eV \\ 
            &  & $N_f$ = (4.68$\pm$3.32)$\times$10$^{17}$ eV$^{-1}$cm$^{-3}$ \\ \hline
    USD-R03 & $\varphi_h$ = (0.285$\pm$0.005) eV & $\varphi_e$ = (0.28$\pm$0.00) eV \\
            & $N_f$ = (4.34$\pm$0.11)$\times$10$^{18}$ eV$^{-1}$cm$^{-3}$ & $N_f$ = (1.83$\pm$0.33)$\times$10$^{18}$ eV$^{-1}$cm$^{-3}$ \\ \hline
    USD-W03 & $\varphi_h$ = 0.31 eV & $\varphi_e$ = (0.295$\pm$0.005) eV \\ 
            & $\varphi_e$ = (0.31$\pm$0.00) eV & $\varphi_h$ = (0.29$\pm$0.00) eV \\
            & $N_f$ = (2.25$\pm$0.02)$\times$10$^{18}$ eV$^{-1}$cm$^{-3}$ & $N_f$ = (1.94$\pm$0.04)$\times$10$^{18}$ eV$^{-1}$cm$^{-3}$ \\ \hline
  \end{tabular}
  \label{t:results}
\end{table*}
\section{Discussion and Prediction}
\label{sec:pred}
\begin{figure} [htbp]
  \centering
  \includegraphics[clip,width=0.9\linewidth]{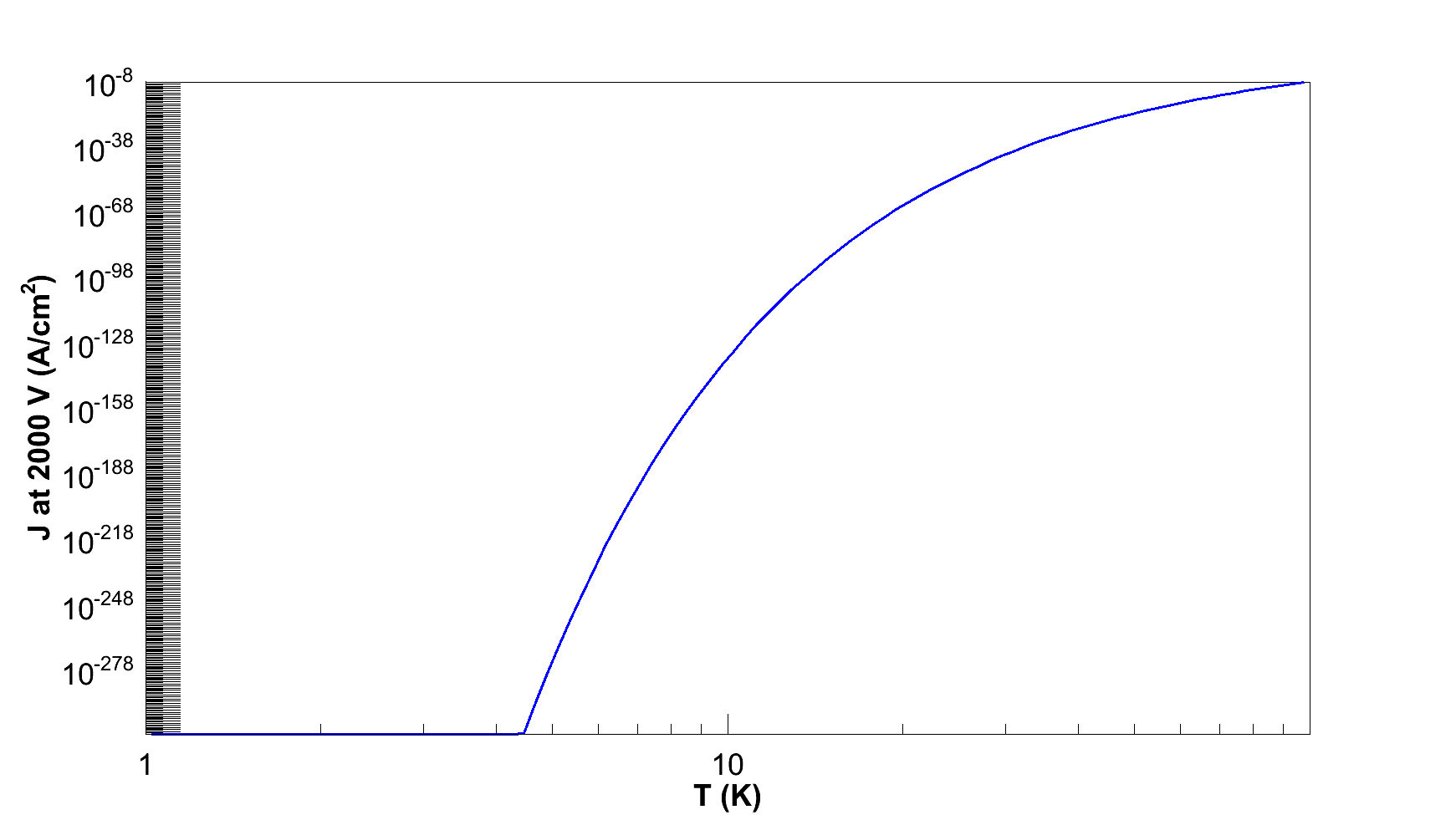}
  \caption{Calculated center contact leakage current density from hole injection at the top contact and electron injection at the bottom contact of the detector, USD-R03, at a negative bias voltage of 2000 V with the configuration shown in Fig.~\ref{fig:cryo}. The leakage current density was estimated based on the contact model parameters extracted from the fits to the measured data shown in Fig.~\ref{fig:h90} and Fig.~\ref{fig:e90}.}
  \label{fig:JT_planar}
\end{figure}

\begin{figure} [htbp]
  \centering
  \includegraphics[clip,width=0.9\linewidth]{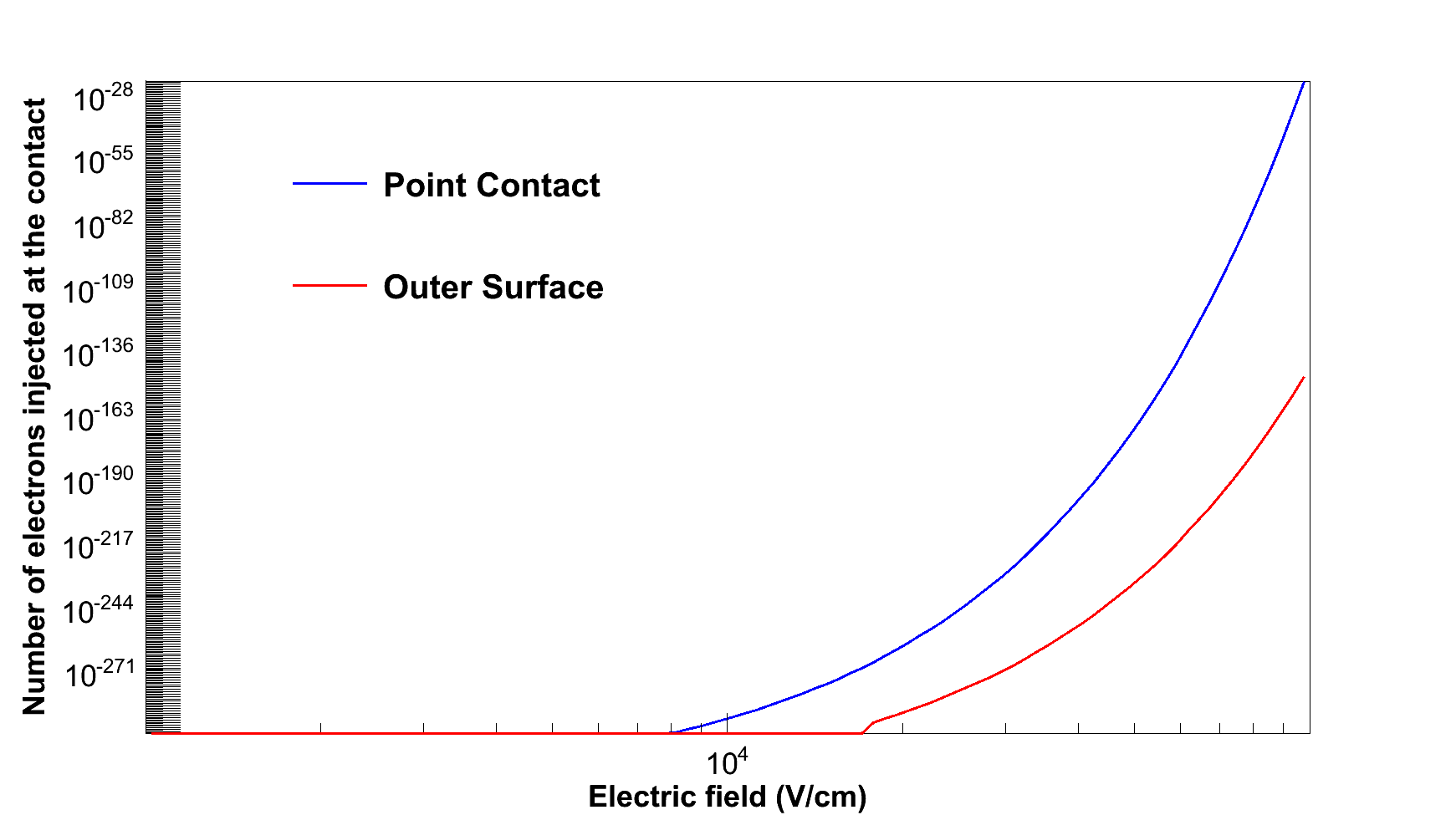}
  \caption{Calculated number of electrons injected to both point contact and outer surface contact as a function of the electric field at 4 K for a PPC Ge detector. The size of this PPC detector is assumed to be 7 cm in diameter and 5 cm in height, corresponding to about 1.02 kg of mass. The point contact area is assumed to be 0.01 cm$^2$. The number of electrons was estimated based on the leakage current density shown in Fig.~\ref{fig:JT_planar} assuming one year of data taking.}
  \label{fig:eE}
\end{figure}

Since the fabrication of Ge detectors with a-Ge contacts has been demonstrated to be consistent and reliable at USD, we can predicate the bulk leakage at liquid helium temperature ($\sim$4 K) assuming that the charge barrier height and the density of states near the Fermi energy level are independent of temperature below 77 K. As an example shown in Fig.~\ref{fig:JT_planar}, the parameters of the barrier height and the density of localized energy states for each contact have been used to predict the leakage as a function of temperature using the the D$\ddot{o}$hler-Brodsky model. The result shows that the bulk leakage quickly approaches nearly zero when temperature approaches 4 K. Based on the leakage current density predicted in Fig.~\ref{fig:JT_planar}, we can also predict the number of electrons injected at the contacts as a function of the electrical field at 4 K for a PPC Ge detector with 7 cm in diameter and 5 cm in height. The area of the point contact is assumed to be 0.01 cm$^2$ and the data-taking time is assumed to be one year (see Fig.~\ref{fig:eE}).

As can be seen in Fig.~\ref{fig:eE}, the number of charge carriers injected from the contacts are so small (nearly 0) for a PPC detector at a high field up to 10$^5$ V/cm. Note that the charge barrier height of the a-Ge is caused by the difference in the band structure between the amorphous semiconductor and the crystalline semiconductor in the interface region. Thus, the electric field intensity in the internal amplification region will not directly affect the charge barrier height of the a-Ge layer. However, the barrier height lowing term is a function of bias voltage, as described in Eq.~\ref{e:j1}. The sum of the barrier height and the barrier lowing term gives an effective barrier height at a given bias voltage. This barrier height lowering term is also correlated to the density of localized energy states near the Fermi level. For a given bias voltage of 2000 volts, as shown in Fig.~\ref{fig:JT_planar}, the barrier lowering terms was taken into account when the leakage current is calculated for USD-R03. The value of the barrier height lowering term is 0.003 eV for the top contact and 0.005 eV for the bottom contact, which are much smaller than the barrier height of 0.285 eV for the top contact and 0.28 eV for the bottom contact.

\section{Conclusions}
\label{sec:conl}
Three guard-ring detectors fabricated from USD-grown crystals are used to measure the bulk leakage current. By applying a well-understood Richardson constant into the D$\ddot{o}$hler-Brodsky model, we have determined the charge barrier height for electrons and holes separately for the top and bottom contacts of three detectors. The density of localized energy states near the Fermi level has also been obtained for the top and bottom contacts of the three detectors. We conclude that the bulk leakage current at helium temperature is negligible, as shown in Fig.~\ref{fig:JT_planar}. 
As we stated earlier, the thermal emission in the bulk of Ge detector is extremely low at $\sim$4 K, corresponding to a negligible level. The surface leakage is inversely proportional to the resistivity of a-Ge, corresponding to a level of $\sim$pA at nitrogen temperature.  Using a guard-ring structure, the surface leakage current can be separated from the bulk leakage current. Since the surface leakage current has a strong dependence on temperature due to the resistance increases as temperature decreases, the surface leakage current is expected to be nearly zero at $\sim$4 k. Thus, the surface leakage will not contribute to the charge read-out from the central contact. Therefore, we conclude that the Ge detectors with a-Ge contacts possess the potential to achieve the sensitivity to measure a single e-h pair, which can be created by extremely low-energy recoils induced by LDM in terms of an extremely small bulk leakage current at liquid helium temperature, as shown in Fig.~\ref{fig:eE}.  

\begin{acknowledgement}
The authors would like to thank Mark Amman for his instructions on fabricating planar detectors and Christina Keller for a careful reading of this manuscript. We would also like to thank the Nuclear Science Division at Lawrence Berkeley National Laboratory for providing us a testing cryostat. This work was supported in part by NSF OISE 1743790, NSF PHYS 1902577, NSF OIA 1738695, DOE grant DE-FG02-10ER46709, DE-SC0004768, the Office of Research at the University of South Dakota and a research center supported by the State of South Dakota.
\end{acknowledgement}


\begin{thebibliography}{99}
\bibitem{ess2012} R. Essig, J. Mardon and T. Volansky, \emph{Direct detection of sub-GeV dark matter}, \emph{Phys. Rev. D} {\bf 85} (2012) 076007.
\bibitem{ess2016} R. Essig et al., \emph{Direct detection of sub-GeV dark matter with semiconductor targets},  \emph{J. High Energ. Phys.} {\bf 2016} (2016) 46. 
\bibitem{ho} C.M. Ho and R.J. Scherrer, \emph{Limits on MeV dark matter from the effective number of neutrinos}, \emph{Phys. Rev. D} {\bf 87} (2013) 023505.
\bibitem{ste} G. Steigman, \emph{Equivalent neutrinos, light WIMPs, and the chimera of dark radiation}, \emph{Phys. Rev. D} {\bf 87} (2013) 103517.  
\bibitem{cd09} Z. Ahmed et al. (CDMS Collaboration), \emph{Search for Weakly Interacting Massive Particles with the First Five-Tower Data from the Cryogenic Dark Matter Search at the Soudan Underground Laboratory}, \emph{Phys. Rev. Lett.} {\bf 102} (2009) 011301.
\bibitem{cd} S.K. Liu et al. (CDEX Collaboration), \emph{Limits on light WIMPs with a germanium detector at 177 eVee threshold at the China Jinping Underground Laboratory}, \emph{Phys. Rev. D} {\bf 90} (2014) 032003.
\bibitem{cd1} H. Jiang  et al. (CDEX Collaboration), \emph{Limits on Light Weakly Interacting Massive Particles from the First 102.8 kg $\times$ day Data of the CDEX-10 Experiment}, \emph{Phys. Rev. Lett.} {\bf 120} (2018) 241301.
\bibitem{cd2} L. T. Yang  et al. (CDEX Collaboration), \emph{Search for Light Weakly-Interacting-Massive-Particle Dark Matter by Annual Modulation Analysis with a Point-Contact Germanium Detector at the China Jinping Underground Laboratory}, \emph{Phys. Rev. Lett.} {\bf 123} (2019), 221301.
\bibitem{cog} Aalseth et al. (CoGeNT Collaboration), \emph{Experimental constraints on a dark matter origin for the DAMA annual modulation effect}, \emph{Phys. Rev. Lett.} {\bf 101} (2008) 251301.
\bibitem{cre12} G. Angloher et al. (CRESST Collaboration), \emph{Results from 730 kg days of the CRESST-II Dark Matter Search}, \emph{Eur. Phys. J. C} {\bf 72} (2012) 1971.
\bibitem{cou} E. Behnke et al. (COUPP Collaboration), \emph{First dark matter search results from a 4-kg CF$_{3}$I bubble chamber operated in a deep underground site}, \emph{Phys. Rev. D} {\bf 86} (2012) 052001. 
\bibitem{bar} J. Barreto et al., \emph{Direct Search for Low Mass Dark Matter Particles with CCDs}, \emph{Phys. Lett. B} {\bf 711} (2012) 264.
\bibitem{dama} R. Bernabei et al. (DAMA/LIBRA Collaboration), \emph{New results from DAMA/LIBRA}, \emph{Eur. Phys. J. C} {\bf 67} (2010) 39.
\bibitem{dar} M. Bossa et al. (DarkSide Collaboration), \emph{DarkSide-50, a background free experiment for dark matter searches}, \emph{JINST} {\bf 9} (2014) C01034.
\bibitem{dri} J.B.R. Battat et al. (DRIFT Collaboration), \emph{First background-free limit from a directional dark matter experiment: results from a fully fiducialised DRIFT detector}, \emph{Phys. Dark Univ.} {\bf 9} (2014) 1.
\bibitem{ede} E. Armengaud et al. (EDELWEISS Collaboration), \emph{Background studies for the EDELWEISS dark matter experiment}, \emph{Astropart. Phys.} {\bf 47} (2013) 1.
\bibitem{kim} S.C. Kim et al. (KIMS Collaboration), \emph{The recent results from KIMS experiment}, \emph{J. Phys. Conf. Ser.} {\bf 384} (2012) 012020.
\bibitem{lux} D. S. Akerib et al. (LUX Collaboration), \emph{Results from a search for dark matter in the complete LUX exposure}, \emph{Phys. Rev. Lett.} {\bf 118} (2017) 021303.
\bibitem{pan} M. Xiao et al. (PandaX Collaboration), \emph{First dark matter search results from the PandaX-I experiment}, \emph{Sci. China Phys. Mech. Astron.} {\bf 57} (2014) 2024.
\bibitem{pic} C. Amole et al. (PICO Collaboration), \emph{Dark Matter Search Results from the PICO-2L C$_{3}$F$_{8}$ Bubble Chamber}, \emph{Phys. Rev. Lett.} {\bf 114} (2015) 231302.
\bibitem{cd14} R. Agnese et al. (SuperCDMS Collaboration), \emph{Search for Low-Mass WIMPs with SuperCDMS}, \emph{Phys. Rev. Lett.} {\bf 112} (2014) 241302.
\bibitem{xe11} E. Aprile et al. (XENON Collaboration), \emph{Design and Performance of the XENON10 Dark Matter Experiment}, \emph{Astropart. Phys.} {\bf 34} (2011) 679. 
\bibitem{xe15} E. Aprile et al. (XENON Collaboration), \emph{Exclusion of leptophilic dark matter models using XENON100 electronic recoil data}, \emph{Science} {\bf 349} (2015) 851.
\bibitem{xe17} E. Aprile et al. (XENON Collaboration), \emph{First Dark Matter Search Results from the XENON1T Experiment}, \emph{Phys. Rev. Lett.} {\bf119} (2017) 181301.
\bibitem{xma} K. Abe et al. (XMASS Collaboration), \emph{Distillation of Liquid Xenon to Remove Krypton}, \emph{Astropart. Phys.} {\bf 31} (2009) 290.
\bibitem{zep} D.Y. Akimov et al. (ZEPLIN-III Collaboration), \emph{Limits on inelastic dark matter from ZEPLIN-III}, \emph{Phys. Lett. B} {\bf 692} (2010) 180.
\bibitem{kad} F. Kadribasic et al., \emph{Directional Sensitivity In Light-Mass Dark Matter Searches With Single-Electron Resolution Ionization Detectors}, \emph{Phys. Rev. Lett.} {\bf 120} (2018) 111301. 
\bibitem{cre17} G. Angloher et al., \emph{Results on MeV-scale dark matter from a gram-scale cryogenic calorimeter operated above ground}, \emph{Eur. Phys. J. C} {\bf 77} (2017) 637.
\bibitem{dam17} A. Aguilar-Arevalo et al. (DAMIC Collaboration), \emph{First Direct-Detection Constraints on eV-Scale Hidden-Photon Dark Matter with DAMIC at SNOLAB}, \emph{Phys. Rev. Lett.} {\bf 118} (2017) 141803.
\bibitem{ahl} S. Ahlen et al., \emph{Limits on cold dark matter candidates from an ultralow background germanium spectrometer}, \emph{Phys. Lett. B} {\bf 195} (1987) 603.
\bibitem{mei} D.-M. Mei et al., \emph{Direct detection of MeV-scale dark matter utilizing germanium internal amplification for the charge created by the ionization of impurities}, \emph{Eur. Phys. J. C} {\bf 78} (2018) 187.
\bibitem{looker} Q. Looker, M. Amman and K. Vetter, \emph{Leakage current in high-purity germanium detectors with amorphous semiconductor contacts}, \emph{Nucl. Instr. and Meth. A} {\bf 777} (2015) 138-147.
\bibitem{luke3} P.N. Luke et al., \emph{Germanium orthogonal strip detectors with amorphous-semiconductor contacts}, \emph{IEEE Trans. Nucl. Sci.} {\bf 47} (no.4) (2000) 1360-1363.
\bibitem{amman1} M. Amman and P.N. Luke, \emph{Position-sensitive germanium detectors for gamma-ray
imaging and spectroscopy}, \emph{Proc. SPIE} {\bf 2} (4141) (2000) 144.
\bibitem{amman2} M. Amman and P.N. Luke, \emph{Three-dimensional position sensing and field shaping in
orthogonal-strip germanium gamma-ray detectors}, \emph{Nucl. Instr. and Meth. A} {\bf 452} (2000) 155-166.
\bibitem{luke1} P.N. Luke et al., \emph{Amorphous Ge bipolar blocking contacts on Ge detectors}, \emph{IEEE Trans. Nucl. Sci.} {\bf NS-39} (4) (1992) 590.
\bibitem{luke2} P.N. Luke, R.H. Pehl and F.A. Dilmanian, \emph{A 140-element Ge detector fabricated with
amorphous Ge blocking contacts}, \emph{IEEE Trans. Nucl. Sci.} {\bf NS-41} (1994) 976.
\bibitem{hull} E.L. Hull and R.H. Pehl, \emph{Amorphous germanium contacts on germanium detectors}, \emph{Nucl. Instr. and Meth. A} {\bf 538} (2005) 651-656.
\bibitem{luke4} P.N. Luke, C.S. Tindall and M. Amman, \emph{Proximity charge sensing with semiconductor detectors}, \emph{IEEE Trans. Nucl. Sci.} {\bf 56} (no.3) (2009) 808-812.
\bibitem{amman3} M. Amman et al., \emph{Proximity electrode signal readout of high-purity Ge detectors}, \emph{IEEE Trans. Nucl. Sci.} {\bf 60} (2013) 1213.
\bibitem{amman4} M. Amman, \emph{Optimization of Amorphous Germanium Electrical Contacts and Surface Coatings on High Purity Germanium Radiation Detectors}, arXiv:1809.03046.
\bibitem{wei} W.-Z. Wei et al., \emph{Investigation of Amorphous Germanium Contact Properties with Planar Detectors Made from Home-Grown Germanium Crystals}, \emph{JINST} {\bf 13} (2018) P12026.
\bibitem{meng} X.-H. Meng et al., \emph{Fabrication and characterization of high-purity germanium detectors with amorphous germanium contacts}, \emph{JINST} {\bf 14} (2019) P02019.
\bibitem{amman5} M. Amman, P.N. Luke, S.E. Boggs, \emph{Amorphous-semiconductor-contact germanium-based detectors gamma-ray imaging and spectroscopy}, \emph{Nucl. Instr. and Meth. A} {\bf 579} (2007) 886-890.
\bibitem{dmwz} D.-M. Mei and W.-Z. Wei, Physics Letters B 785 (2018) 610-614.
\bibitem{barton} P. Barton, M. Amman, R. Martin, and K. Vetter, "Ultra-low noise mechanically cooled germanium detector", Nucl. Instr. Meth. in Phys. Res.  A, 812 (2016) 17-23. 
\bibitem{han} W.L. Hansen and E.E. Haller, \emph{Amorphous germanium as an electron or hole blocking contact on high-purity germanium detectors}, \emph{IEEE Trans. Nucl. Sci.} {\bf NS-24} (1) (1977) 61.
\bibitem{pru} S. Prussin, D.I. Margolese and R.N. Tauber, \emph{Formation of amorphous layers by ion
implantation}, \emph{Journal of Applied Physics} {\bf 57} (1985) 180-185.
\bibitem{mot1} N. F. Mott, "Conduction in Non-crystalline Materials," Philosophical Magazine, vol. 19,
pp. 835-852, 1969.
\bibitem{mot2} N. F. Mott and E. A. Davis, Electronic Processes in Non-Crystalline Materials, Oxford:
Clarendon Press, 1971.
\bibitem{brod2} M.H. Brodsky and G.H. D$\ddot{o}$hler, \emph{A new type of junction: amorphous/crystalline}, \emph{Crit. Rev. Solid State Mater. Sci.} {\bf 5} (no.4) (1975) 591-595.
\bibitem{sze} S.M. Sze, \emph{Physics of Semiconductor Devices}, \emph{Wiley, New York}, 1981.
\bibitem{zeg} B. Van Zeghbroeck, 2011, http://ece-www.colorado.edu/~bart/book/.
\bibitem{hhl} Hanhui Liu, et al., \emph{Ohmic contact formation of metal/amorphous-Ge/n-Ge junctions with an anomalous modulation of Schottky barrier height}, \emph{Applied Physics Letters} {\bf 105}, 192103, 2014.
\bibitem{dohl} G.H. D$\ddot{o}$hler and M.H. Brodsky, \emph{Amorphous-crystalline heterojunctions}, \emph{Proceedings International Conference Tetrahedrally Bonded Amorphous Semiconductors}, 1974, p.351.
\bibitem{brod1} M.H. Brodsky, G.H.D$\ddot{o}$hler and P.J. Steinhardt, \emph{On the measurement of the conductivity density of states of evaporated amorphous silicon films}, \emph{Phys. Stat. Sol.} {\bf 72} (1975) 761-769.
\bibitem{heni} H.K. Henisch, \emph{Rectifying Semi-Conductor Contacts}, Oxford University Press, Oxford, 1957.
\end{thebibliography}
\end{document}